\documentclass[prb,amsmath,amssymb,floatfix,twocolumn,citeautoscript]{revtex4}
\usepackage{bm}
\usepackage{epsfig}

\begin{document}

\title{Inelastic Electron Tunneling Spectroscopy for Topological Insulators}

\author{Jian-Huang She$^1$, Jonas Fransson$^2$, A. R. Bishop$^1$, Alexander V. Balatsky$^{1,3}$}

\affiliation{$^1$Theoretical Division, Los Alamos National Laboratory, Los Alamos, NM, 87545, USA.\\
$^2$Department of Physics and Astronomy, Uppsala University, Box 530, SE-751 21 Uppsala, Sweden.\\
$^3$Center for Integrated Nanotechnologies, Los Alamos National Laboratory, Los Alamos, NM, 87545, USA.}

\begin{abstract}

Inelastic Electron Tuneling Spectroscopy (IETS) is a powerful spectroscopy that allows one to investigate the nature of local excitations and energy transfer in the system of interest. We study IETS for Topological Insulators (TI) and investigate the role of inelastic scattering on the Dirac node states on the surface of TIs. Local inelastic scattering is shown to significantly modify the Dirac node spectrum. In the weak coupling limit, peaks and steps are induced in second derivative $d^2I/dV^2$. In the strong coupling limit, the local negative U centers are formed at impurity sites, and the Dirac cone structure is fully destroyed locally. At intermediate coupling resonance peaks emerge. We map out the evolution of the resonance peaks from weak to strong coupling, which interpolate nicely between the two limits. There is a sudden qualitative change of behavior at intermediate coupling, indicating the possible existence of a local quantum phase transition. We also find that even for a simple local phonon mode the inherent coupling of spin and orbital degrees in TI leads to the spin polarized texture in inelastic Friedel oscillations induced by local mode. 

\end{abstract}

\date{\today \ [file: \jobname]}

 \maketitle

\textbf{\emph{ Introduction:}} Topological insulators (TIs) are a new state of quantum matter, which is insulating in the bulk but possesses metallic surface states [\onlinecite{Hasan10,Qi11}]. One of the defining properties of TIs is their stability to disorder. There exists a topological invariant associated with the bulk band structure that protects TIs against any time-reversal invarant perturbations. Much attention has been focusing on impurity scattering on the surface of TIs, and a number of novel effects have been revealed, e.g. suppression of backscattering [\onlinecite{Ando98}], absence of Anderson localization [\onlinecite{Bardarson07,Nomura07}], weak antilocalization [\onlinecite{Suzuura02,McCann06,Lu11,Tkachov11}], new quasiparticle interference patterns [\onlinecite{Hu09,Lee09,Franz10,QWang10,JWang11,Liu12}], and impurity-induced resonance states [\onlinecite{Biswas10,Schaffer11,Schaffer12,Kapitulnik12}]. However the scattering processes considered so far are mostly elastic scatterings. Much less effort has been devoted to the study of inelastic scattering processes that involve not only momentum but also energy transfer. Inelastic scattering processes are ultimately related to strong correlation effects, which is of particular theoretical interest in the study of topological matter.

In this paper, we consider the inelastic scattering of Dirac fermions off a single localized vibrational mode on the surface of TIs. The local mode is an impurity with internal degrees of freedom, similar to a magnetic impurity. The crucial difference with a magnetic impurity is that here time reversal symmetry is preserved. A related model of a Fermi gas interacting strongly with a localized phonon was proposed by Yu and Anderson [\onlinecite{Yu84}], where it was found that the Fermi gas softens the phonon frequency and generates an effective double well potential for phonons at sufficiently strong coupling. This model has been heavily investigated subsequently (see e.g. [\onlinecite{Sassetti87,Katsnelson88,Dora07,Dora09}]). 

The local electronic structure is also substantially modified by the presence of the local mode, which can be detected by inelastic electron tunneling spectroscopy (IETS) with a scanning tunneling microscope (STM) [\onlinecite{Ho98,Ho01}]. The pre-STM IETS using metal-insulator-metal tunneling junctions measures the vibrational spectrum of the dynamical defect on the insulator, with the peak in $d^2V/dI^2$ corresponding to the vibrational frequency  [\onlinecite{Lambe66,Lambe68,Scalapino67,Hansma77}]. IETS for s-wave superconductors was considered in [\onlinecite{Brandt70}], and for d-wave superconductors in [\onlinecite{Balatsky03,Morr03,Morr05}], where it was found that at weak coupling there is a kink in the $dI/dV$ curve [\onlinecite{Balatsky03}], and at stronger coupling resonance peaks emerge [\onlinecite{Morr03}]. The interference of incoming and outgoing waves in inelastic scattering can also produce standing wave patterns, i.e. inelastic Friedel oscillations [\onlinecite{Fransson07,Gawronski11}].

In this paper, we calculate the local density of states (LDOS) for Dirac fermions in the presence of 
a local mode. The main results are as follows: (i) In the weak coupling limit, by using second-order perturbation theory, we find that at the impurity site, there is a kink and a logarithmic singularity at the local mode frequency $\omega_0$ in the fermion LDOS. Correspondingly in the frequency derivative of LDOS, i.e. $d^2I/d V^2$, there is a step and a peak at $\omega_0$. Away from the impurity site, inelastic Friedel oscillations are shown to be present in $d^2I/dV^2$. Similar features are found for spin polarized STM. (ii) In the strong coupling limit, where we have a single site problem, the LDOS becomes a series of delta functions, forming a single band at negative frequency. The density of states at the Fermi level is fully depleted. The result here is universal in the sense that it is independent of the band structure of conduction electrons. (iii) By summing over the noncrossing diagrams, we also map out the evolution of LDOS from weak to strong coupling. At intermediate coupling two resonance peaks appear. As coupling further increases, the peak at positive frequency first becomes higher than the one at negative frequency, and then the two peaks change role, with the negative frequency peak higher than the positive frequency one. At even stronger coupling, the positive frequency peak finally dies out.

\textbf{\emph{ Inelastic scattering with a localized mode:}} The low energy Hamiltonian describing the massless Dirac fermions coupled to the local boson at ${\bm r}=0$ is
\begin{equation}
{\cal H}=\sum_{{\bm k}\sigma}\epsilon_{\bm k}c^{\dagger}_{{\bm k}\sigma}c_{{\bm k}\sigma}+\omega_0b^{\dagger}_0b_0+g\sum_{{\bm k}{\bm k}'\sigma}c^{\dagger}_{{\bm k}\sigma}c_{{\bm k}'\sigma}(b^{\dagger}_0+b_0),
\end{equation}
with $\epsilon_{\bm k}=v_f{\hat z}\cdot({\bm \sigma}\times {\bm k})$, where $v_f$ is the Fermi velocity, ${\bm \sigma}=(\sigma_x,\sigma_y,\sigma_z)$ is the vector of Pauli matrices. The bare fermion Green's function $G_0({\bm k},i\omega_n)=1/[i\omega_n-v_f(\sigma_xk_y-\sigma_yk_x)]$ reads in real space and real frequency
\begin{multline}
G_0({\bm r}_1,{\bm r}_2;\omega)=\frac{1}{4v_f^2} \left[ \omega Y_0\left(\frac{|\omega|}{v_f}r \right) -i|\omega|J_0\left(\frac{|\omega|}{v_f}r \right)\right]\\ -{\hat {\bm z}}\cdot({\hat {\bm r}}\times{\bm \sigma})\frac{1}{4v_f^2}\left[\omega J_1\left(\frac{|\omega|}{v_f}r \right) +i|\omega| Y_1\left(\frac{|\omega|}{v_f}r \right) \right],
\end{multline}
where $J_i$ and $Y_i$ are the Bessel function of the first and second kind. The local bosonic mode has propagator ${\cal D}({\bm k},i\omega_n)={\cal D}(i\omega_n)=2\omega_0/[(i\omega_n)^2-\omega_0^2]$.

\begin{figure}
\begin{centering}
\includegraphics[width=0.45\linewidth]{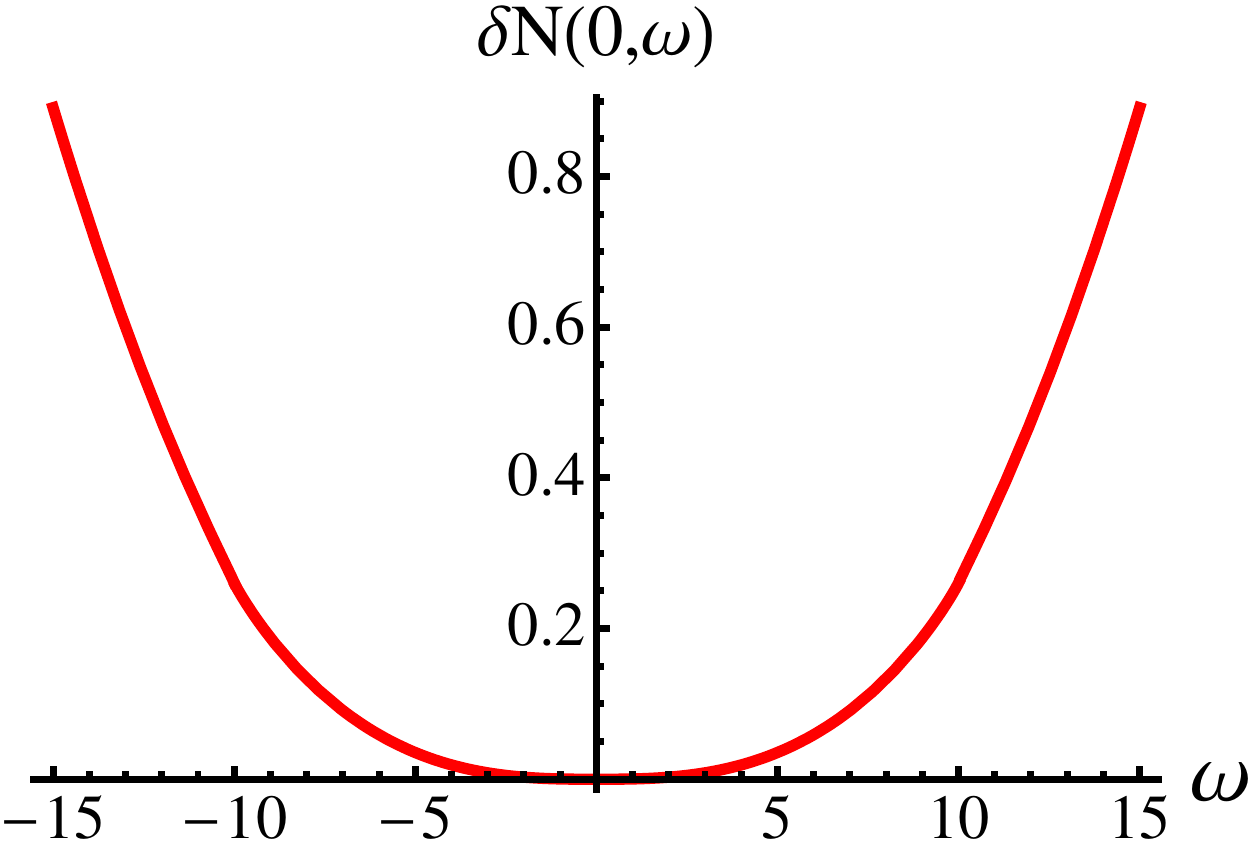} 
\includegraphics[width=0.45\linewidth]{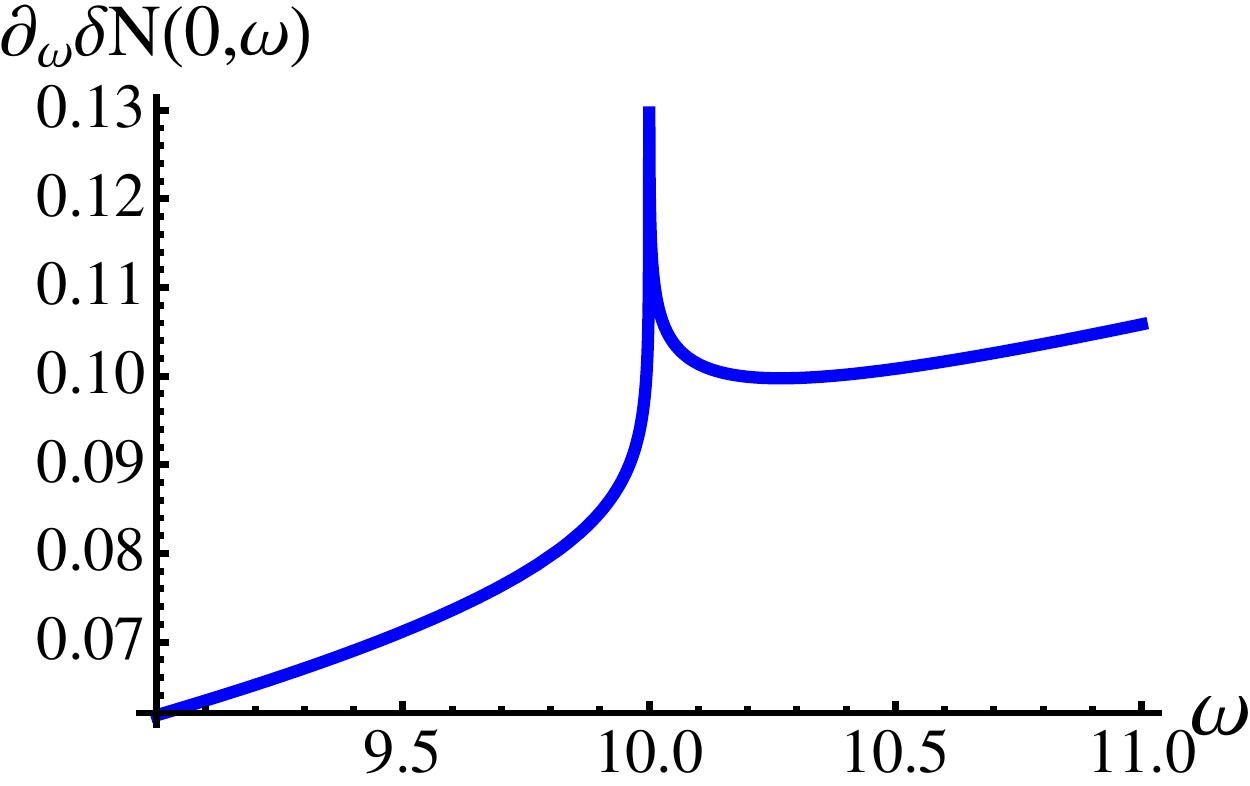} 
\end{centering}
\caption{Change of LDOS and its frequency derivative at the impurity site due to the presence of a local vibrational mode in the weak coupling limit. Here $T= 0$, frequency is measured in units of meV, with local mode frequency $\omega_0=10{\rm meV}$, and the cutoff $\Lambda=300{\rm meV}$.}
\label{LDOS-TI}
\end{figure}

\begin{figure}
\begin{centering}
\includegraphics[width=0.99\linewidth]{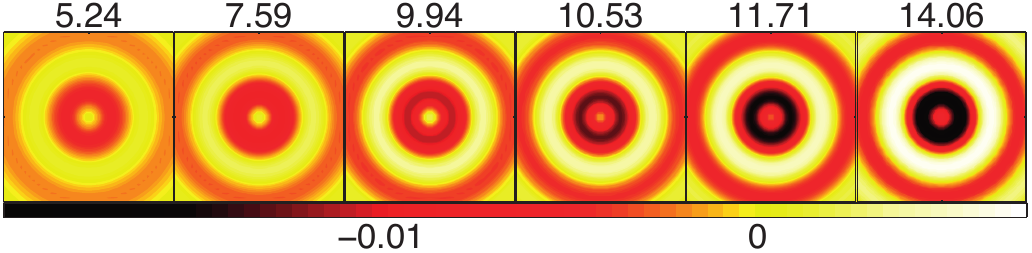}
\includegraphics[width=0.99\linewidth]{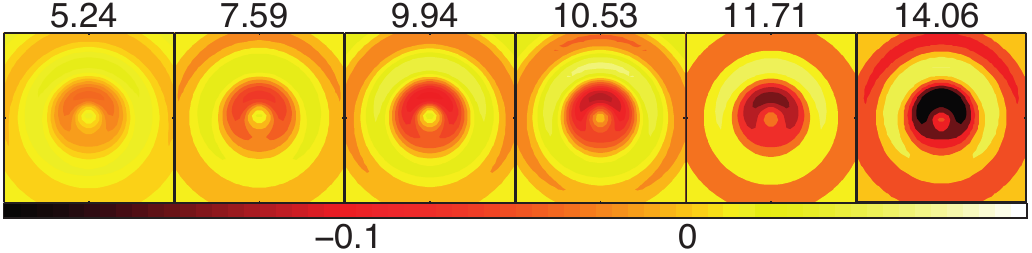}
\end{centering}
\caption{IETS maps of $\partial^2\delta I({\bm r},V)/\partial V^2$ at the voltages (mV) indicated  with $T=1$ K, for spin unpolarized STM (up) and for SP-STM with spin-polarization given by ${\bm m}_0=3n(1,0,0)/4$ (down). }
\label{fig-ddN}
\end{figure}

Multiple scattering of Dirac fermions off the local bosonic mode can be accounted for by the T-matrix, which modifies the fermion Green's function to $G({\bm r},{\bm r}';i\omega_n)=G_0({\bm r},{\bm r}';i\omega_n)+G_0({\bm r},0;i\omega_n){\cal T}(i\omega_n)G_0(0,{\bm r}';i\omega_n)$. The T-matrix is of the form ${\cal T}(i\omega)=\Sigma(i\omega_n)/[1-G_0(0,0;i\omega_n)\Sigma(i\omega_n)]$, where the local self energy is calculated from the one particle irreducible (1PI) diagrams, with result $\Sigma(i\omega_n)=-g^2T\sum_{m} G(0,0;i\omega_n-i\Omega_m){\cal D}(i\Omega_m)$. 

\textbf{\emph{Weak coupling limit:}} When the coupling between the fermions and the local mode is weak, perturbation theory can be employed. In this case, the change of LDOS is small. However, sharp features can be detected in the frequency derivative of LDOS.

For weak scattering, the T-matrix can be approximated by the one loop self energy, $\Sigma^{(0)}(i\omega_n)=-g^2T\sum_{m} G_0(0,0;i\omega_n-i\Omega_m){\cal D}(i\Omega_m)$,
 which reads at zero temperature
\begin{multline}
\Sigma(\omega)=- \frac{\pi g^2}{v_f^2}\left\{(\omega-\omega_0)\ln \left(\frac{\Lambda}{|\omega-\omega_0|}\right)+(\omega_0\to-\omega_0)\right.\\ \left.+i\pi\left[ (\omega-\omega_0)\Theta(\omega-\omega_0)+(\omega\to-\omega) \right]\right\},
\label{sigma0}
\end{multline}
where $\Lambda$ is a cutoff.
At high temperatures, the imaginary part of the self energy at zero frequency becomes linear in temperature, i.e. ${\rm Im}{\Sigma}(\omega=0,T)\simeq -(2\pi^2g^2/v_f^2)T$. It is interesting to compare this result with the effect of inelastic scattering in normal Fermi liquids, where the self energy has $T^2$ dependence due to Pauli blocking in the absence of a finite volume Fermi surface.

For Dirac fermions coupled to a localized magnetic atom via exchange interaction $J\sum{\bf S}\cdot c^{\dagger}_{{\bm k}\sigma}{\bm \sigma}c_{{\bm k}'\sigma}$, the T-matrix becomes $g^2T\sum G_0({\bm k},i\omega_n-i\Omega_m)\sigma_i\sigma_j\chi_{ij}(i\Omega_m)$, where $\chi_{ij}$ is the local spin susceptibility. We notice that, for an isotropic $\chi_{ij}$, the effect on the LDOS will be the same as that of a local phonon at least for weak coupling.

The effect of such inelastic scattering processes can be detected by STM, where LDOS $N({\bm r},\omega)=-(1/\pi){\rm Tr}\,{\rm Im}\,G({\bm r},{\bm r};\omega)$ is measured. The change of LDOS has two contributions, one is of the form $\delta N^{(1)}\sim {\rm Re}\left[{\rm Tr}G_0^2({\bm r},0;\omega)\right]{\rm Im}{\Sigma}(\omega)$, the other reads $\delta N^{(2)}\sim {\rm Im}\left[{\rm Tr}G_0^2({\bm r},0;\omega)\right]{\rm Re}\Sigma(\omega)$. 

Consider first LDOS at the site of the local vibrational mode. Noticing that $G_0(0,0;\omega)=\frac{1}{4v_f^2}\left(\frac{2\omega}{\pi}\ln \frac{|\omega|}{\Lambda}-i|\omega| \right)$, for $\omega>0, T=0$, one obtains
\begin{eqnarray}
\delta N^{(1)}=\lambda\omega^2{\cal C}_1(\omega-\omega_0)\Theta(\omega-\omega_0),\\
\delta N^{(2)}=-\lambda \omega^2{\cal C}_2 \left[(\omega-\omega_0)\ln \frac{\Lambda}{|\omega-\omega_0|}+(\omega_0\to -\omega_0)\right],
\end{eqnarray}
where ${\cal C}_1=\pi[(4/\pi^2)\ln^2(\omega/\Lambda)-1 ]$, ${\cal C}_2=(4/\pi)\ln(\omega/\Lambda)$, and the overall coefficient $\lambda\propto g^2$. Such a  signal can be seen most clearly from the frequency derivative of LDOS, i.e. $d^2I/dV^2$ (see Fig.~\ref{LDOS-TI}). Only above the threshold of the boson frequency can the local mode be excited, scattering over which renders the fermion lifetime finite. Thus ${\rm Im}\Sigma(\omega)$ is nonzero only for $\omega>\omega_0$. This gives rise to a kink at $\omega=\omega_0$ in $\delta N^{(1)}$, or equivalently a step in $\partial_{\omega}\delta N^{(1)}$. Such a sharp feature also enters ${\rm Re}\Sigma(\omega)$ via the  Kramers-Kronig relation, resulting in a peak in $\partial_{\omega}\delta N^{(2)}$.

The effect of inelastic scattering will be present even away from the scattering center. Interference between incoming and outgoing Dirac fermions produce standing waves on the surface of a TI, similar but qualitatively different from conventional Friedel oscillations. These standing waves can be seen in the oscillations of the second derivative $\partial^2I({\bm r},V)/\partial V^2$, in the narrow window of energies near the energy of the bosonic mode $\omega_0$ where inelastic scattering occurs. 

From $\delta N({\bm r},\omega)\sim G_0^2({\bm r},0;\omega)\Sigma(\omega)$, one can see that the spatial dependence of the change of LDOS comes from the free fermion Green's function $G_0({\bm r},0;\omega)$. From the asymptotic behavior of the Bessel functions, $\delta N({\bm r},\omega)\sim (1/r)\cos(2\omega r/v_f +\theta_0)$, the period of the standing waves is set by the bias and the Fermi velocity. For the on-site LDOS, the sharp features, which have their origin from $\Sigma(\omega)$, occur near the boson frequency $\omega_0$. We plot in Fig.~\ref{fig-ddN} the spatial variation of $\partial_\omega\delta N({\bm r},\omega)$ at $T=1$ K. The main change in this variation as one sweeps through the inelastic mode is the center peak (below $\omega_0$) which turns into a dip (above $\omega_0$). 

A salient feature of the surface state of topological insulators is the strong coupling between spin and orbital degrees of freedom, resulting in the locking of spin dynamics to that of charge. Such a feature is also manifest in inelastic scattering processes. Even scattering off a non-magnetic mode produces sharp features in the magnetic structure ${\bm M}({\bm r},\omega)=-{\rm Tr}\,{\rm Im}\, G({\bm r},{\bm r};\omega)\boldsymbol{\sigma}/(2\pi)$, which can be detected by spin-polarized STM (SP-STM) with a magnetic tip, distinguishing itself completely from a normal metal, for which ${\bm M}({\bm r},\omega)$ vanishes identically in the presence of non-magnetic impurities.

The second order change of ${\bm M}({\bm r},\omega)$ due to the presence of a local mode is $\delta{\bm M}({\bm r},\omega)\sim{\rm Im}g_0\Sigma g_\perp[\hat{\bm r}\times\hat{\bm z}]$, where $G_0({\bf r},\omega)=g_0({\bm r},\omega)\sigma^0+g_\perp({\bm r},\omega)[\hat{\bm r}\times\hat{\bf z}]\cdot\boldsymbol{\sigma}$. One can see immediately that $\delta M_z=0$, i.e. the induced magnetic structure is in-plane only. We plot in Fig.~\ref{fig-ddN} the spatial variation of $\partial_\omega\delta {\bm M}({\bm r},\omega)$ at different frequencies, with the magnetic direction of the tip parallel to the surface. One can clearly see anisotropic inelastic Friedel oscillations near the boson frequency $\omega_0$.

\begin{figure}
\begin{centering}
\includegraphics[width=0.45\linewidth]{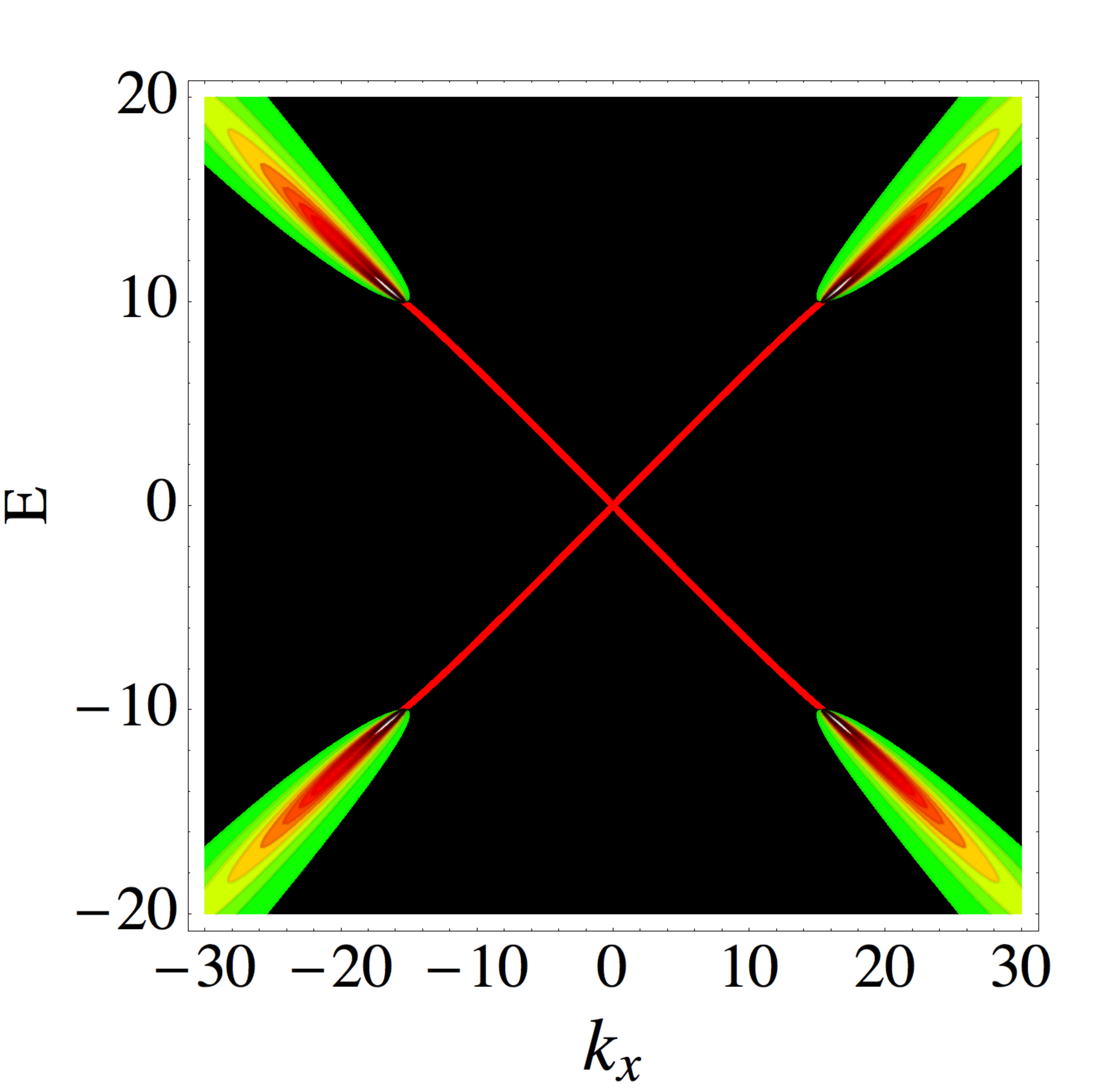} 
\includegraphics[width=0.45\linewidth]{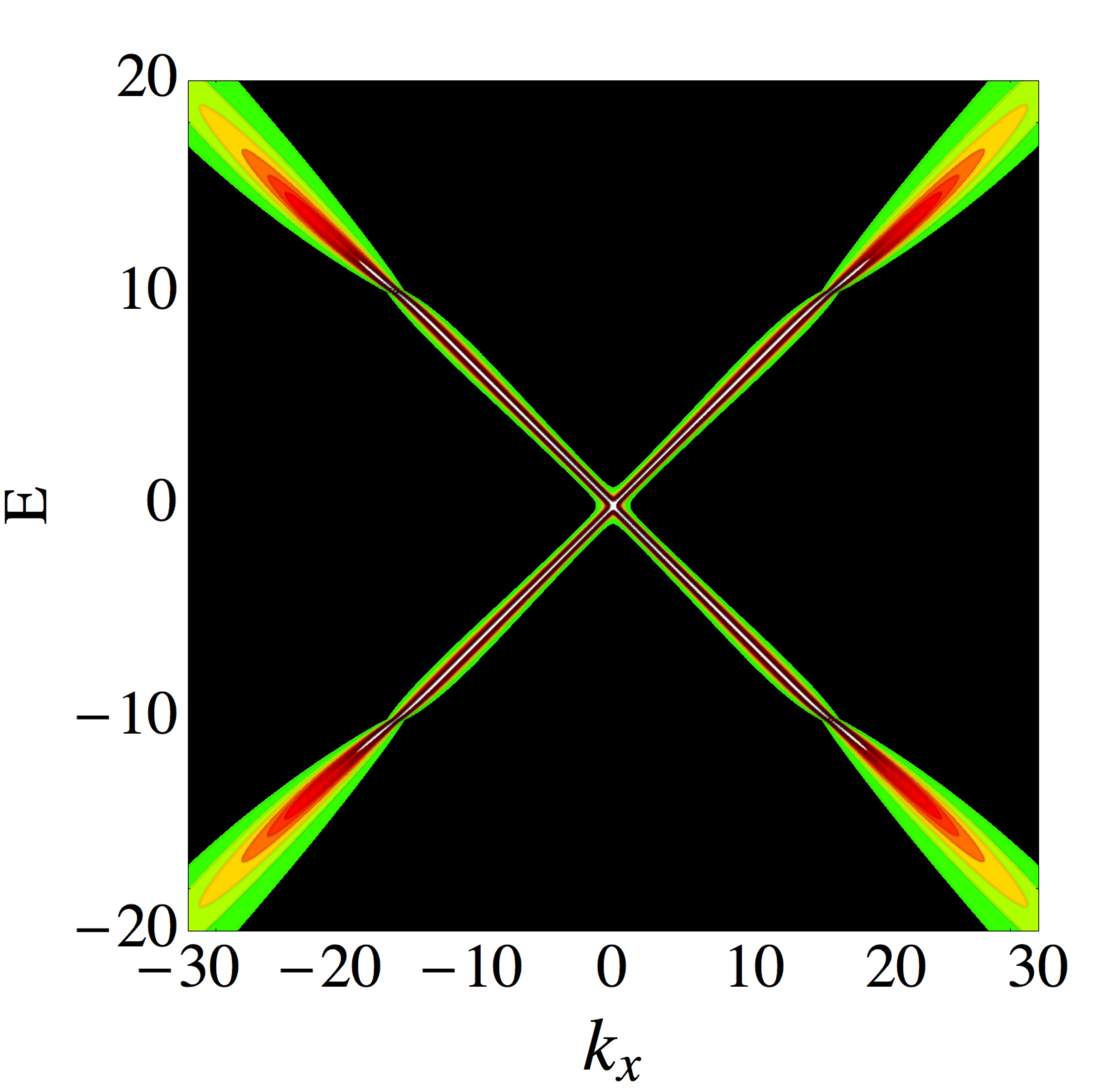} 
\end{centering}
\caption{Impurity-averaged spectral function at $T=0, 23{\rm K}$ for $k_y=0$, with $\omega_0=10{\rm meV}, \Lambda=300{\rm meV}, \pi g^2n_i/v_f^2=0.1$.}
\label{Akw-TI}
\end{figure}

Besides local probes, e.g. STM, the effect of inelastic scattering on the surface of TI can also be measured by more averaging probes, e.g. ARPES, transport. To illustrate this, we consider a finite but low density of localized vibrational modes on the surface of a TI. Up to first order in the impurity density $n_{\rm imp}$, the impurity-averaged fermion Green's function reads ${\hat G}({\bm k},\omega)=G_0({\bm k},\omega)\left[1-{\hat\Sigma}(\omega)G_0({\bm k},\omega)\right]^{-1}
$, where the self energy in the Born approximation is ${\hat\Sigma}(\omega)=n_i{\cal T}(\omega)\simeq n_i{\Sigma}(\omega)$. The resulting spectral function $A_{\sigma}({\bm k},\omega)=-(1/\pi){\rm Im}{\hat G}_{\sigma\sigma}({\bm k},\omega)$ at $T=0$ is plotted in Fig.~\ref{Akw-TI}.  For $|\omega|<\omega_0$, since ${\rm Im}\,{\hat\Sigma}(\omega)=0$, the spectral function is not smeared by inelastic processes.\footnote{In real materials, the spectral function is broadened by other processes, e.g. elastic scattering from point impurities and edge impurities.} Only the dispersion is slightly modified, with $A({\bm k},\omega)=(1/2)[ \delta(\omega-{\rm Re}\,{\hat\Sigma}(\omega)+v_Fk)+ \delta(\omega-{\rm Re}{\hat\Sigma}(\omega)-v_Fk)]$, retaining the Dirac cone structure. For $|\omega|>\omega_0$, inelastic scattering gives rise to finite ${\rm Im}\,{\hat\Sigma}(\omega)$, and the spectral function is broadened. There is a qualitative change at $|\omega|=\omega_0$. Although dressed by thermal fluctuations, one can see from Fig.~\ref{Akw-TI} that such a pattern is still visible at relatively high temperatures.

\textbf{\emph{Strong coupling limit:}} We consider here the strong coupling limit, where two fermions with opposite spins are trapped on the impurity site. In this case, one only needs to consider the single site problem with coupled fermion and boson. This is a standard textbook problem (see e.g. [\onlinecite{Mahan}]). However the problem here has a crucial difference with the independent boson model solved in [\onlinecite{Mahan}]. Here since the effective onsite energy for fermions is negative, in the ground state the fermionic state is doubly occupied.

As $g\to \infty$, the model reduces to a single site problem, $H=\sum_{\sigma}\epsilon_{\sigma}c^{\dagger}_{\sigma}c_{\sigma}+\omega_0b^{\dagger}b+g\sum_{\sigma}c^{\dagger}_{\sigma}c_{\sigma}(b^{\dagger}+b)$, which can be exactly diagonalized by making the transformation ${\tilde H}=e^sHe^{-s}$, with 
 $s=\frac{g}{\omega_0}\sum_{\sigma}c^{\dagger}_{\sigma}c_{\sigma}(b^{\dagger}-b)$, and the resulting Hamiltonian reads 
\begin{equation}
{\tilde H}=\omega_0b^{\dagger}b+\sum_{\sigma}\left(\epsilon_{\sigma}-\frac{g^2}{\omega_0}\right)c^{\dagger}_{\sigma}c_{\sigma}-\frac{2g^2}{\omega_0}c^{\dagger}_{\uparrow}c_{\uparrow}c^{\dagger}_{\downarrow}c_{\downarrow}.
\end{equation}
The last term represents an attractive Hubbard interaction generated by electron-phonon coupling. The local mode thus forms a negative-U center.

With $\epsilon_{\sigma}=0$, in the ground state, two electrons are trapped at the impurity site, while the phonon occupation is zero. We also notice that, in the strong coupling limit, the effective on-site energy $\epsilon_{\sigma}-g^2/\omega_0$ is always negative, independent of the band structure. The real-time electron Green's function $G_{\sigma\sigma'}(t)=-i\langle Tc_{\sigma}(t)c_{\sigma'}^{\dagger}(0)\rangle$ can be written in terms of the transformed quantities, $G_{\sigma\sigma'}(t)=i\theta(-t) e^{\beta\Omega}{\rm Tr}\left( e^{-\beta {\tilde H}}{\tilde c}^{\dagger}_{\sigma'}e^{i{\tilde H}t} {\tilde c}_{\sigma}e^{-i{\tilde H}t} \right)$. Here ${\tilde c}_{\sigma}=c_{\sigma}X$ and $ {\tilde c}^{\dagger}_{\sigma'}=c^{\dagger}_{\sigma'} X^{\dagger}$, and the operator $X=\exp\left[ -\frac{g}{\omega_0}(b^{\dagger}-b) \right]$. Since the transformed Hamiltonian is a summation of the electron part and the phonon part, ${\tilde H}={\tilde H}_c+{\tilde H}_b$, the Green's function can be factorized as
\begin{multline}
G_{\sigma\sigma'}(t)=i\theta(-t)e^{\beta\Omega_c}{\rm Tr}\left[e^{-\beta{\tilde H}_c}c_{\sigma'}^{\dagger} e^{i{\tilde H}_ct} c_{\sigma}e^{-i{\tilde H}_ct} \right]\\e^{\beta\Omega_b} {\rm Tr}\left[e^{-\beta{\tilde H}_b}X^{\dagger}(0)X(t)\right],
\end{multline}
The phonon part gives contribution $e^{- \Phi(t)}$, with $\Phi(t)=\frac{g^2}{\omega_0^2}\left(  1-e^{i\omega_0 t}\right)$ at zero temperature. The electronic part leads to $\exp\left[ -it\left(  \epsilon_\sigma-3g^2/\omega_0\right) \right]\delta_{\sigma\sigma'}$. The resulting spectral function is a series of delta functions forming a single band,
\begin{equation}
A(\omega)= e^{-\lambda}\sum_{n=0}^{\infty} \frac{\lambda^n}{n!}\delta\left(\omega-\epsilon_\sigma+\frac{3g^2}{\omega_0}+n\omega_0\right),
\end{equation}
with $\lambda=g^2/\omega_0^2$.

\begin{figure}
\begin{centering}
\includegraphics[width=0.48\linewidth]{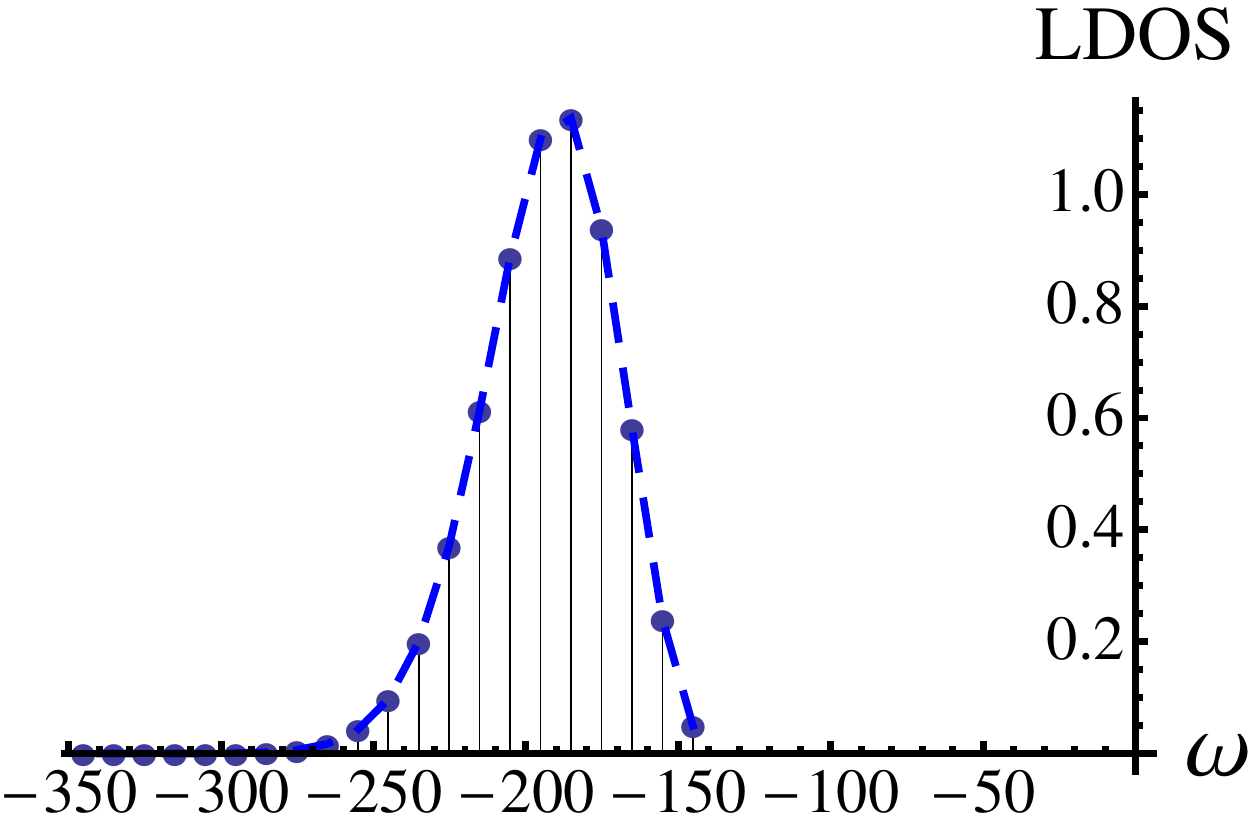} 
\includegraphics[width=0.48\linewidth]{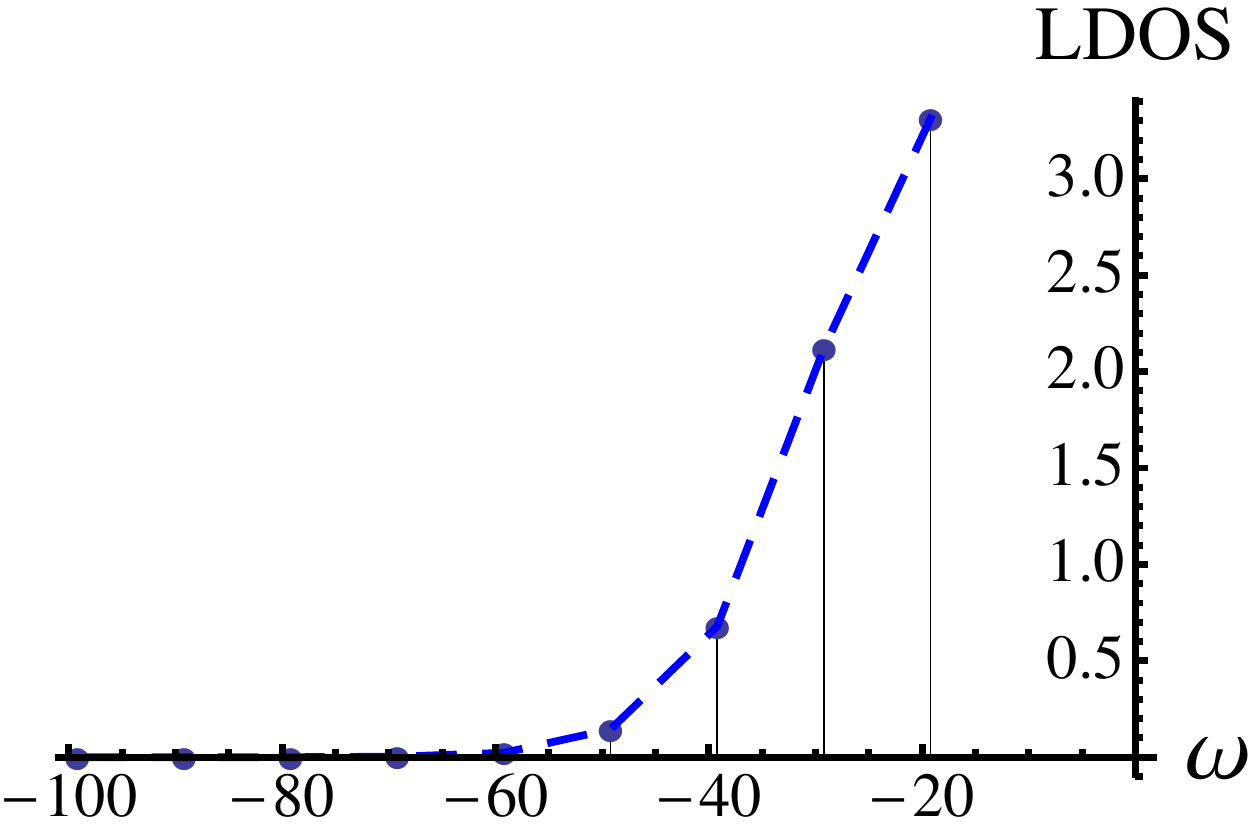} 
\end{centering}
\caption{LDOS at the impurity site in the strong coupling limit, with $g=22$ (left) and $g=8$ (right). Here $\omega_0=10{\rm meV}$. }
\label{LDOS-Iboson}
\end{figure}

\textbf{\emph{Evolution from weak to strong coupling:}}  Having considered two limiting cases with weak and strong coupling, now we proceed to study generic coupling strength, where  further approximations are required, and the LDOS can only be calculated numerically. We will use the self consistent Born approximation, where the T-matrix obeys the integral equation, 
\begin{equation}
{\cal T}^{-1}_n=\left[\Sigma^{(0)}_n+\sum_m{\cal K}_{nm}{\cal T}_m\right]^{-1}-G^{(0)}_n,
\end{equation}
where $G^{(0)}_n\equiv G_0(0,0;i\omega_n)=(-i\omega_n/4\pi v_f^2)\ln(\Lambda^2/\omega_n^2)$, and ${\cal K}_{nm}=-g^2T{\cal D}(i\omega_n-i\nu_m)\left[G_0(0,0;i\nu_m)\right]^2$.

\begin{figure}
\begin{centering}
\includegraphics[width=0.45\linewidth]{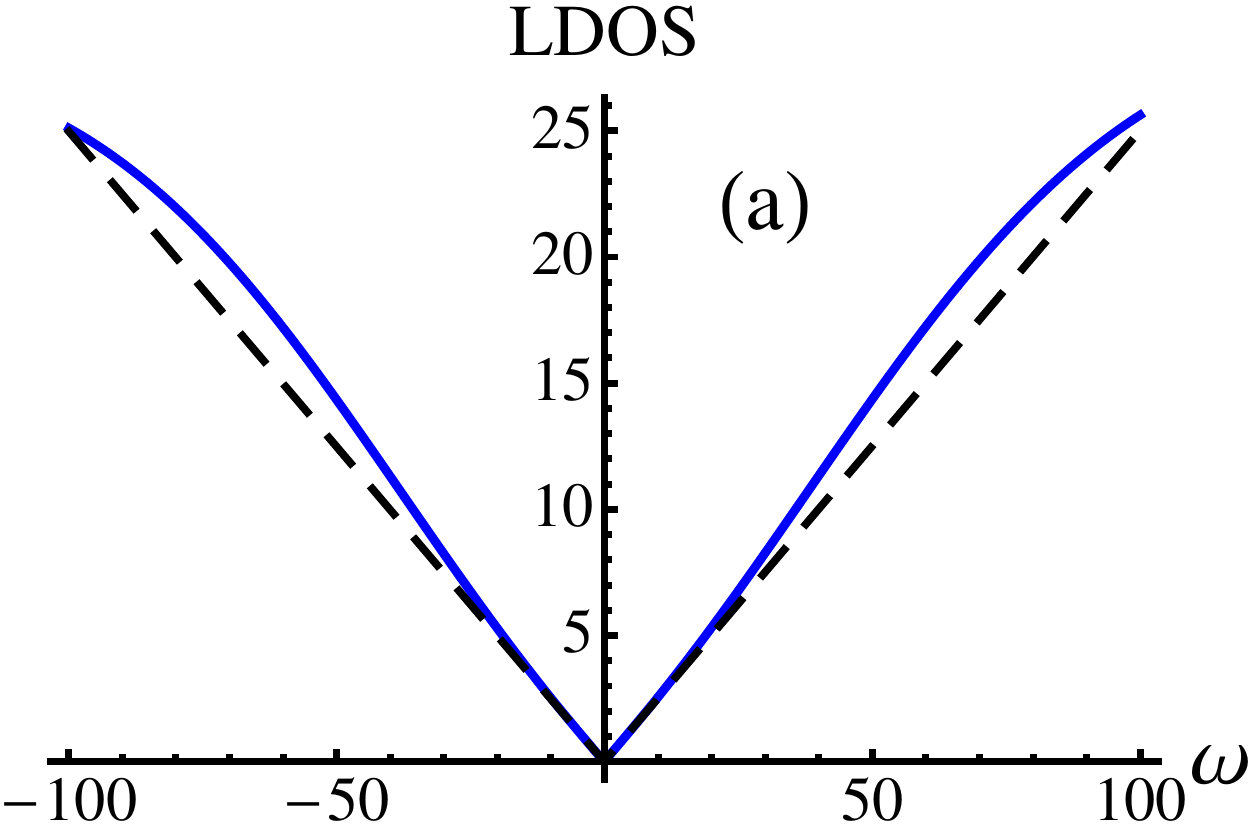} 
\includegraphics[width=0.45\linewidth]{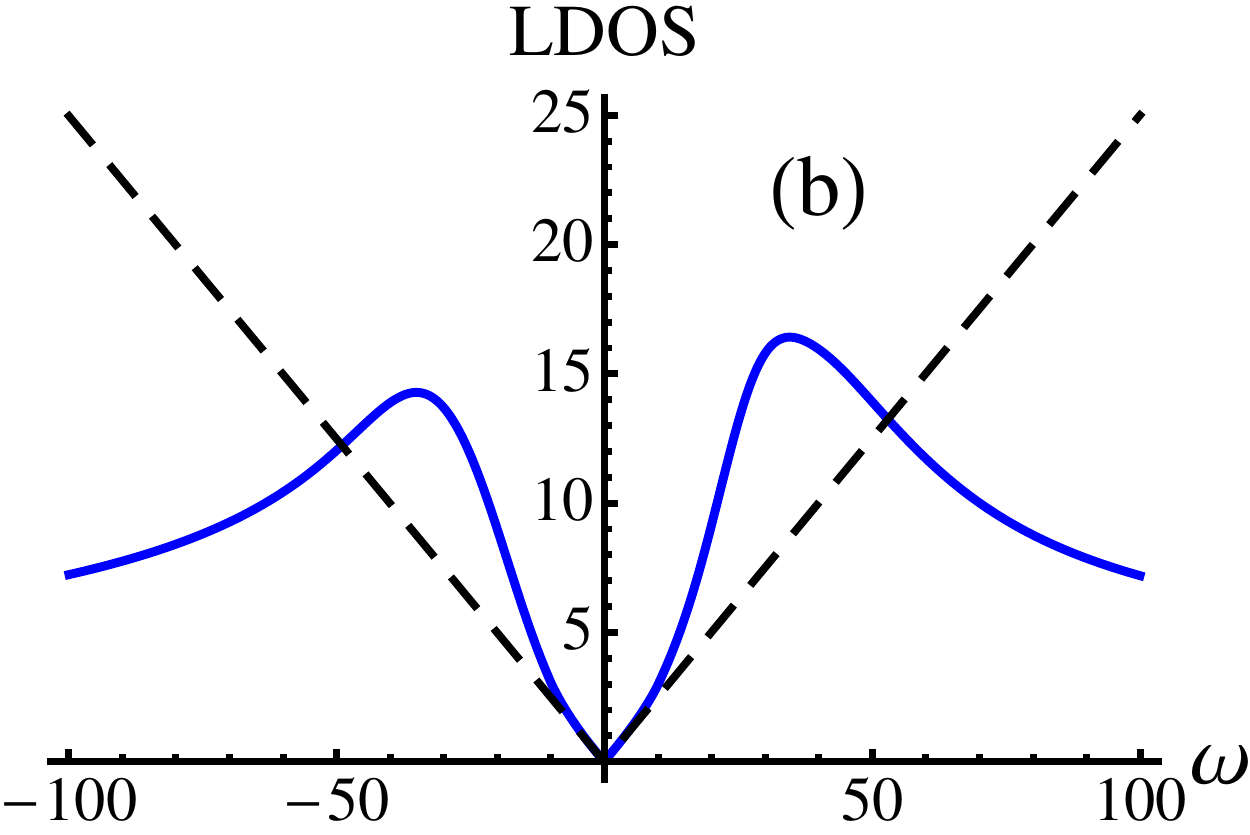} 
\includegraphics[width=0.45\linewidth]{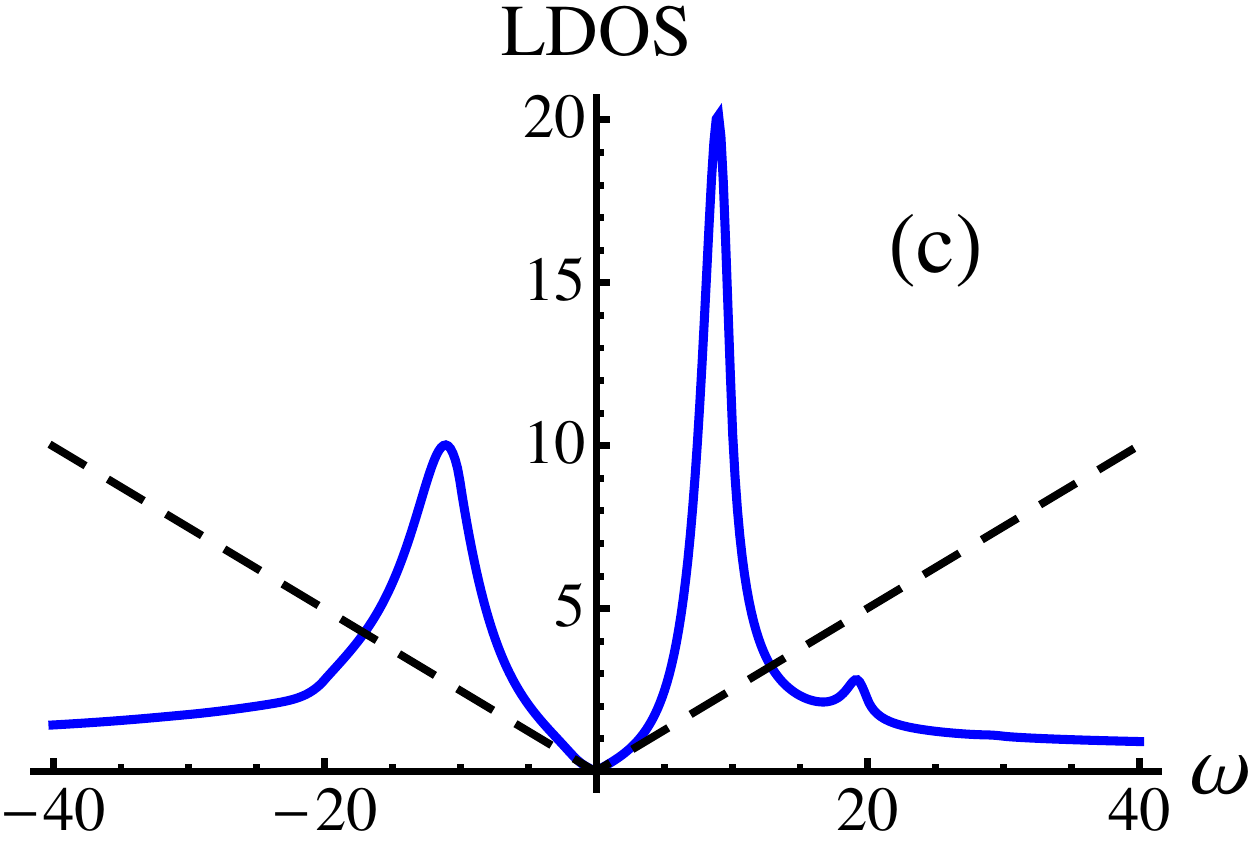} 
\includegraphics[width=0.45\linewidth]{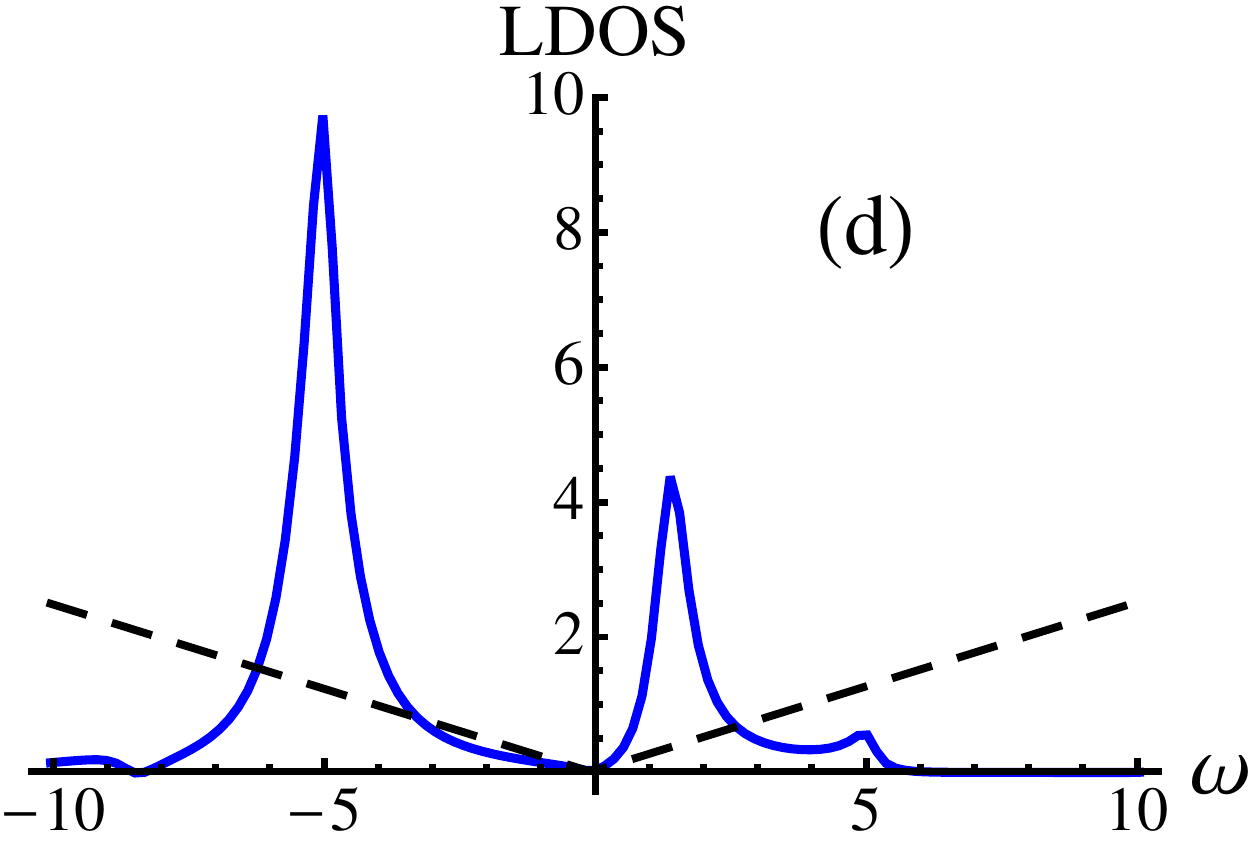} 
\includegraphics[width=0.45\linewidth]{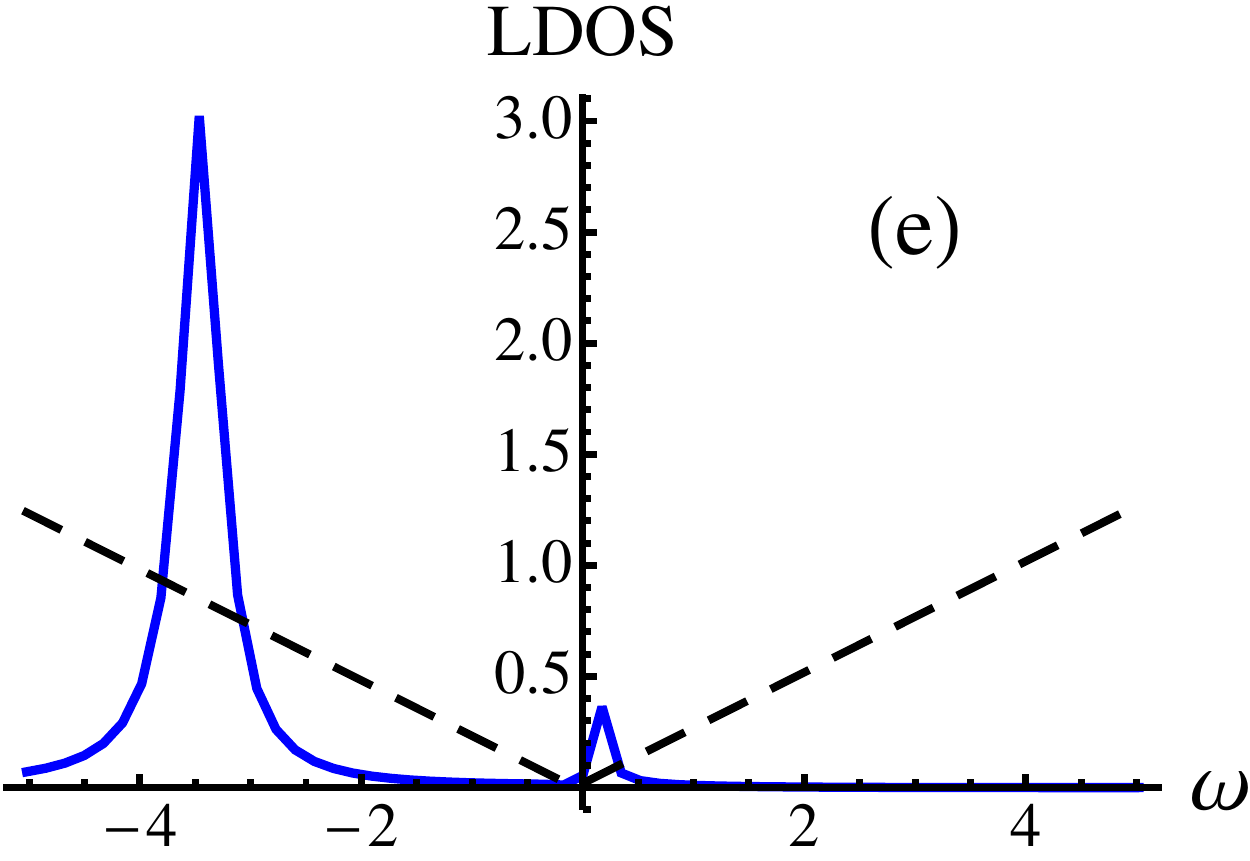} 
\includegraphics[width=0.45\linewidth]{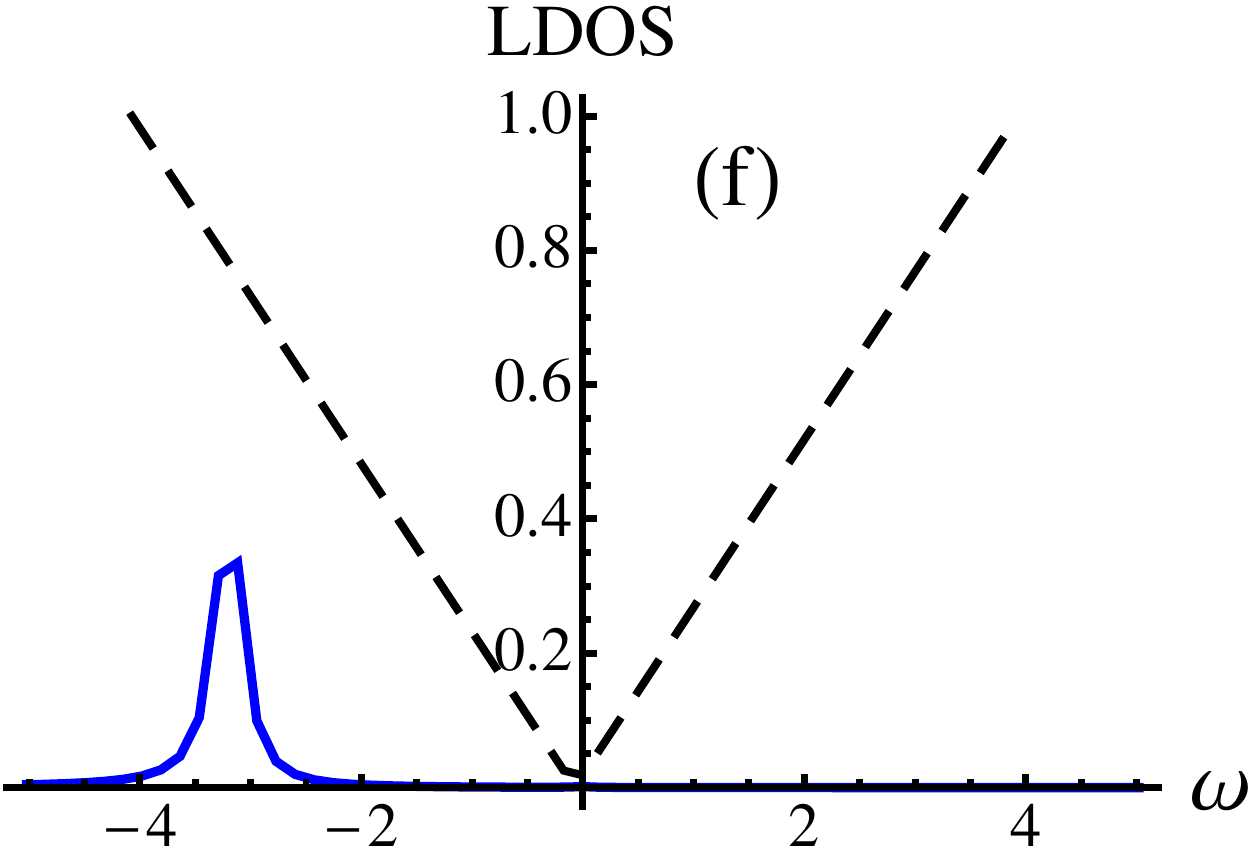} 
\end{centering}
\caption{(Color online) Evolution of LDOS from weak to strong coupling. Here $\omega_0=10 {\rm meV}, \Lambda=300{\rm meV}, v_f=1$, and $g=0.003, 0.01, 0.03, 0.1, 0.3, 1$. }
\label{LDOS3g}
\end{figure}

The numerical results for $T=0$ are shown in Fig~(\ref{LDOS3g}) for various coupling strengths. As we increase coupling, the LDOS starts to deviate from the Dirac cone strucutre at $\omega \gtrsim\omega_0$ (Fig~\ref{LDOS3g}a), with the spectral weight accumulating at intermediate frequencies. Two resonance peaks form as coupling further increases (Fig~\ref{LDOS3g}b). At still  stronger coupling (Fig~\ref{LDOS3g}c), the two resonance peaks move towards lower frequency, with the peak height increasing sharply, the peak width becoming much narrower, and the asymmetry in two peaks  more pronounced.

We notice that in  Fig~(\ref{LDOS3g}a,b,c), the linear dispersion is retained near $\omega=0$. The LDOS is modified only at larger frequencies. However, qualitative changes occur at even stronger coupling. While from Fig~\ref{LDOS3g}b to Fig~\ref{LDOS3g}c, more and more spectral weight moves into the peak at positive frequency, in Fig~\ref{LDOS3g}d, the peak at negative frequency becomes the dominant one. With increasing coupling, the peak at positive frequency is further suppressed (Fig~\ref{LDOS3g}e), and finally becomes invisible (Fig~\ref{LDOS3g}f), as expected from the above result in the strong coupling limit. The LDOS at negative frequency also gets depleted near $\omega=0$, with the Dirac cone structure fully destroyed (Fig~\ref{LDOS3g}d,e,f).

While the changes from Fig~\ref{LDOS3g}a to Fig~\ref{LDOS3g}c, and from Fig~\ref{LDOS3g}d to Fig~\ref{LDOS3g}f are continuous, there is a sudden qualitative change from Fig~\ref{LDOS3g}c to Fig~\ref{LDOS3g}d, indicating the possible existence of a phase transition \footnote{The numerics become bad for $0.04\lesssim g\lesssim 0.09$, which may result form the critical fluctuations near a phase transition.}. More sophisticated methods, e.g. numerical renormalization group (NRG), are needed to settle this issue. The model considered here can be regarded as a generalization of the extensively studied pseudogap Kondo model (PKM) (see [\onlinecite{Vojta06}] and references therein), with the defects having discrete internal degrees of freedom generalized to defects having continuous internal degrees of freedom. In PKM, where the conduction electrons have a power law DOS $\rho(\epsilon)\sim |\epsilon|^r$, weak coupling renormalization group calculations found an unstable RG fixed point, indicating a phase transition between weak and strong coupling [\onlinecite{Fradkin90}]. However this method is only valid for small $r$, and nonperturbative NRG studies showed that the phase transition disappears for $r\geq 1/2$ in the particle-hole symmetric case [\onlinecite{Ingersent98}].

\textbf{\emph{ Conclusion:}} In conclusion, we have studied the inelastic scattering of Dirac fermions with a single impurity that has continuous internal degrees of freedom and does not break time reserval symmetry. For weak coupling, sharp features near the local mode frequency (step and peak) can be detected in $d^2I/dV^2$ using IETS-STM. At intermediate coupling, resonance peaks appear in $dI/dV$ curve. The original Dirac cone structure is fully destroyed locally at strong coupling. There is possibly a zero temperature phase transition, and thus a local quantum critical state, at intermediate coupling, which we leave for future work. The presence of a finite density of such local modes will substantially change the transport properties of TIs. It would be interesting to explore the dephasing effect of such modes, to study how weak antilocalization is influenced by inelastic scattering.

We acknowledge useful discussions with Tanmoy Das, and Jian-Xin Zhu. This work was supported, in part, by UCOP-TR01, by  the Center for Integrated Nanotechnologies, a U.S. Department of Energy, Office of Basic Energy Sciences user facility. Los Alamos National Laboratory, an affirmative action equal opportunity employer, is operated by Los Alamos National Security, LLC, for the National Nuclear Security Administration of the U.S. Department of Energy under contract DE-AC52-06NA25396. JF acknowledge the support of the Swedish Research Council.

\bibliographystyle{apsrev}
\bibliography{strings,refs}

\end{document}